\documentclass{article}

\usepackage{arxiv}

\usepackage[utf8]{inputenc} 
\usepackage[T1]{fontenc}    
\usepackage{hyperref}       
\usepackage{url}            
\usepackage{booktabs}       
\usepackage{amsfonts}       
\usepackage{nicefrac}       
\usepackage{microtype}      
\usepackage{graphicx}
\usepackage{amsmath}
\usepackage{multirow}

\begin{document}

\title{Pricing  Multivariate European Equity Option Using Gaussians Mixture Distributions and EVT-Based Copulas.}
\author{\texttt{Hassane Abba Mallam}$^{1},$\texttt{\ Diakarya Barro}$^{2}$%
	\AND	\texttt{Yam\'{e}ogo WendKouni}$^{3}$ \texttt{and Bisso Saley%
	}$^{1}.$ \AND $^{1}${\small UAM, Universit\'{e} Abdou Moumouni (Niger)} \AND 
	$^{2}${\small \ Universit\'{e} Thomas Sankara (Burkina Faso). } \AND $^{3}$%
	{\small \ LANIBIO, Universit\'{e} Joseph Ki-Zerbo (Burkina Faso). }}

\maketitle
	\begin{abstract}
		In this article, we present an approach which allows to take into account
		the effect of extreme values in the modeling of financial asset returns and
		in the valorisation of associeted options. Specifically, the marginal
		distribution of assets returns is modeled by a mixture of two gaussiens
		distributions. \ Moreover , we model the joint dependence structure of the
		returnsusing a copula function, the extremal one, which is suitable for our
		financial data, particularly the extreme value copula. Applications are made
		on the Atos and Dassault Systems actions of the CAC40 index. Monte-Carlo
		method is used to compute the values of some equity options such as the call
		on maximum, the call on minimum, the digital option and the spreads option
		with the basket (Atos, Dassault systems) as underlying.

		\bigskip \textbf{Keywords:}  Options, Extremes values, Gaussians mixture; Copulas; Monte-Carlo; %
		European market. 
		
		2010 Mathematical Subject Classification: 60E05; 91B24; 91G20; 91G60; 91G70.%
			
		2010 JEL classification: C02; F31; G13.
	\end{abstract}

\section{Introduction}

Since the pionering work of Black-Scholes \cite {bs1973} and Cox and al. \cite {cox1979} (respectively in the continuous and discrete case), option pricing has become a crucial topic in finance. Indeed, considering an european-type  option on an underlying asset with a price $S_t $; strike $K$ and expiration $T$, Black and Scholes have made it possible to determine a formula for the price of such options under certain assumptions, the fundamental of which are the lack of arbitrage opportunity  and that on the price $S_t$ of the asset underlying ($S_t$ follows geometric Brownian motion), i.e.

\begin{equation}
dS_t=\mu S_tdt + \sigma S_tW_t, 	\nonumber		
\end{equation}

where $\mu$ and $\sigma$ are constant and $W_t$ is a standard geometric Brownian motion.

Thus, the formula of the relative theoretical value $\dfrac{C_t}{S_t}$ of a call option is then given by:
\begin{equation}
C^{all}_t= N(d_1)-\kappa e^{-r(T-t)}N(d_2),
\end{equation}
and the relative theoretical value of a put is given by:
\begin{equation}
P^{ut}_t=-N(-d_1)+\kappa e^{-r(T-t)}N(-d_2),
\end{equation}
where $N(.)$ is the standard normal distribution; $ d_1 = \dfrac {(r+ \frac{\sigma^2}{2}) (T-t) - \ln (k)} {\sigma \sqrt {T-t}} $; $ d_2 = d_1- \sigma \sqrt {T-t} $ and $ \kappa = \frac {K} {S_t} $ the relative strike and $r$ the risk-free interest rate.\newline

Options are essential financial products allowing to their holders to hedge against the risk of falls in their investments. This is how we are increasingly seeing the creation of several types of options such as exotic options; multivariate options; etc., with the aim of providing more security. As a result, valuation models are  also evolving. Of all the multiple option pricing models, it turns out that each one is primarily based on the dynamics of underlying asset  pricing model (for options with only one underlying) or the asset portfolio (for options on multiple assets), when market assumptions are known. In fact, since the assumption of no arbitrage opportunity (NAO) in the markets is the basis of the fundamental results obtained in finance, it's considered by default \footnote {There are markets on which the arbitration assumption is considered.}. The advantage under this NAO assumption is that, associated with that of market completeness, there is a single risk-neutral probability for which the discounted flows are martingales. In the univariate case, one of the most interesting results obtained in this direction on valuation is that of Breeden and al. \cite {bred1978}. It states that the second derivative (when it exists and continuous) of the price of a standard option relatively to the strike coincides with the risk-neutral density. Indeed, if $D_t$ is the price of a european option of an underlying asset with price $X_t$ having for pay-off $g (X_T)$, $T$ the time to expiration and $r$ the risk-free interest rate; then the risk-neutral density $f^* (X_T)$ is linked to $D_t $ by:

\begin{equation}
D_t=e^{-r(T-t)}\mathbb{E}^*\{g(X_T)\}=e^{-r(T-t)}\int g(X_T)f^*(X_T)dX_T.
\end{equation}

For valuation in the multivariate framework, this risk-neutral formula is a simple generalization. Talponen and al. \cite{jarno2014} recently gave the multivariate version of the univariate result. \newline

Multivariate options (rainbow; digital; quantos;etc.), which will be the main subject of our study in this paper, constitute the central themes of current research on financial risk coverage. The advantage lies in the fact that they offer better coverage against risks. Indeed, the basic idea is that when the option is a function of several assets, the fall in value of one asset is compensated by the rise of another asset in the portfolio. Thus the association or dependence between assets plays a major role in the pricing of these types of options. To take such an aspect into account in the valuation, the use of the copula is a good alternative.\newline

The valuation of multivariate options by copulas is in full development. The copula gives the advantage of joining the marginal and the dependence structure. This is the case for many works on the valuation of options with copulas, the emphasis is first on marginal risk-neutral densities and then on the joint risk-neutral density (risk-neutral copula). For example, we can cite the work of, Cherubuni and al. \cite {cheru2000}; Cherubuni and al \cite {cheru2003},  Rosenberg and al. \cite {rosb1999}, Salmon and al. \cite {mark2000}; Slavchev and al. \cite {slavchev2009}. However, all this work did not take into account the effects of extreme values in the marginal, which is not without effect on valuation (risk of over-valuation or under-valuation). However, there are other copula modeling approaches based on volatility dynamics as in Goorbergh and al. \cite{goorbergh2004}. The reader can consult this last one for full details on the literature on this approach.

In this present study, we propose a valuation method for multivariate options allowing to take into account the effects of extreme values in the marginal and the joint structure on the basis of the works of Edier et al \cite{jul2008} and the use of  extreme values copulas.\newline
In the rest of this work, the first section we give the results obtained by Idier et al\cite{jul2008}. which will be necessary and somes essentielles notions on copula. In the second section, we expose the methodology used for leading properly the application of the demarche. Then, the obtained results of different estimations and simulations are presented, with their analysis and interpretations. The last section present a conclusion and discussion.

\section{Preliminaries}

\subsection{Results of an approach of modeling financial assets}

It is prouved that the empirical distribution of financial assets returns has thicker tails than that of the Gaussian distribution. This indicates the presence of extreme values. This fact shows also that the normal distribution does not make it possible to model rigorously the returns of financial assets because it does not take into account the extremes. This is the case with the method proposed by Black-Scholes. \newline

To take into account the effects of extreme values, Idier et al \cite{jul2008} proposed, as an alternative to the normal distribution, to model the distribution of the rates of return of the underlying asset of a univariate option, under NAO assmption, by a mixture of Gaussian distribution in the continuous framework \footnote{Their method is a generalization of the method in the discrete case of Bertholon, Monfort and Pegoraro(2006); Pegoraro(2006)}.
They justified their choice by the fact that a mixture of Gaussian distributions makes it possible to approximate all the distributions usually used (Gaussian; alpha-stable; Student; hyperbolic; etc.); also that it has certain theoretical properties allowing easy handling in the frame a theoretical model for valuing asset price; that it is easy to simulate and can reproduce various sets (mean, variance, skewness and kurtosis) observed in the data.\newline

Under the assumption that the historical distribution of the returns of the underlying $X_ {t + 1} = \ln (\frac {S_ {t + 1}} {S_t})$ where $S_t$ is the price at time $t$ of the underlying asset, is a mixture of $ 2 $ Gaussian distribution. Its density is given by:
\begin{equation}
f(x)=\sum_{i=1}^{2} p_i.n(x, \mu_i, \sigma_i^2),
\end{equation}

where
 $$ n (x, \mu_i, \sigma_i^2) = \frac {1}{\sigma_i \sqrt {2\pi}} \exp \bigg\{\dfrac {-(x-u_i) ^ 2}{2 \sigma_i ^ 2} \bigg\}, $$ is the density of a Gaussian distribution with mean $ \mu_i $ and standard deviation $ \sigma_i $; $ 0 <p_i <1\; and \; \sum_ {i = 1}^{2} p_i = 1.$ \newline

Moreover, the stochastic discount factor is characterized by an affine exponential form, i.e.

\begin{equation}
M_{t; t+1}=\exp\{\alpha_t X_t + \beta_t\}. \label{fact}
\end{equation} 
They establish, under these assumptions, that the risk-neutral distribution is also a Gaussian mixture and that its density $ f ^ {*} $ is defined by:
\begin{equation}
f^{*}(x)= \sum_{i=1}^{2} v_i.n(x, \mu_i +\alpha.\sigma_i^2, \sigma_i^2), \label{risque-neutral density}
\end{equation}

where 
$$v_i=(p_i \exp\{\mu_i \alpha + \frac{\alpha^2}{2}\sigma_i^2\})/( \sum_{i=1}^{2}p_i \exp\{\mu_i \alpha + \frac{\alpha^2}{2}\sigma_i^2) \;  with\;0<v_i<1, \sum_{i=1}^{2} v_i=1.$$

Thus, they derive the relative theoretical price of an European call with a one-period maturity ($ T = t + $ 1) and a relative strike $ k $:
\begin{equation}
c_t(T, k)= \sum_{i=1}^{2} v_i \gamma_i c_{bs}(\sigma^2_i, \frac{k}{\gamma_i}), \label{call idier}
\end{equation}

where $ c_ {bs} (.,.) $ is the Black-Scholes one period $(T=t+1)$ formula for the relative price of a call and $\gamma_i=\exp\{\mu_i +\alpha.\sigma_i^2-r + \frac{\sigma^2_i}{2}  \}$, for $ J = 2 $.

\begin{remark}
		The existence of the call-put parity relation makes it possible to simplify the task in calculating option price. It is then sufficient to calculate the price of the call to deduce that of the corresponding put (or vice versa) by the relation:
	\begin{equation}
	C_t(T,K)+ Ke^{-r(T-t)}= P_t(T, K) + S_t. \label{parite}
	\end{equation} 
\end{remark}

\subsection{A survey of copulas}

In this section we recall the basics notions on copulas. These are the definitions and properties essential for our study. For more details on copulas, see Nelsen \cite{nelsen}.

\subsubsection{Definitions and properties}

The copula is a function allowing to capture the structure of dependence between several random variables. \\

A function $ C: [0,1] ^ {d} \rightarrow \lbrack 0,1] $ is a d-copula if it satisfies the following properties:

\begin{itemize}
	\item[i)]  For all $ u \ in [0,1] $, $ C (1, ... 1, u, 1, ... 1) = u $;
	
	\item[ii)] For all $u_i \ in [0,1] $, $C (u_1, ..., u_d) = 0 $ if at least one of the $u_i$ is zero;
	
	\item[iii)] $C$ is "grounded" and d-increasing, i.e
	
	\begin{equation}
	\sum_{i_{1}=1}^{2}...
	\sum_{i_{d}=1}^{2}(-1)^{i_{1}+...+i_{d}}C(u_{1,i_{1}},...,u_{d,i_{d}}))\geq 0,
	\end{equation}%
	for all $ (u_{1,1},...,u_{d,1})$ and $(u_{1,2},...,u_{d,2})\in \lbrack 0,1]^{d} $ with $u_{d,1}\leq u_{d,2}$.
\end{itemize}

The fundamental result on the copula due to Sklar (1969) states that: For a whole multivariate distribution $ F $ with continuous marginal $ F_ {1} $, ..., $ F_ {d} $, then there exists a unique \footnote {Uniqueness is not guaranteed when marginal are not continuous.} copula $ C: [0,1] ^ {d} \rightarrow \lbrack 0,1] $ such that:

\begin{equation}
F(x_{1},..,x_{d})=C(F_{1}(x_{1}),...,F_{d}(x_{d})). \label{sk}
\end{equation}

Conversely, when $C$ is a copula and $ F_ {1} $, ..., $ F_ {d} $ are marginal distributions, the function $ F $ defined by (\ref{sk}) is a multivariate distribution of marginal distributions $F_ {1}$, ..., $F_ {d}$. \newline

This result makes it possible to deduce several properties of the copula including invariance by any monotonic transformation.
Another consequence of Sklar's theorem is that every copula $ C $ satisfies

\begin{equation}
max\bigg(\sum_{i=1}^d u_i-d+1; 0 \bigg) \leq C(u_1, u_2, ...,u_d) \leq min(u_1, u_2, ..., u_d).
\end{equation}

This relation is the variant in terms of copulas of the Frechet-Hoeffding bounds of a multivariate distribution. The upper bound $ min (u_1, u_2, ..., u_d) $ is the comonotonic copula representing the perfect positive dependence. The lower bound $ max\big(\sum_ {i = 1}^ d u_i-d + 1; 0\big) $ is a copula only for $ d = 2 $. In this case it represents the perfect negative dependence. \newline

\begin{remark}
	If $ \bar{F} $ is the multivariate survival distribution of a $ F $ distribution of marginals $ F_i; \; i = 1, .., d $, then the survival copula, denoted by $ \bar{C} $, is defined by:
	\begin{equation*}
	\bar{F}(x_1,...,x_d)=\bar{C}(\bar{F}_1(x_1),...,\bar{F}_d(x_d)).
	\end{equation*}
	The survival copula $ \bar{C} $ is related to the copula $ C $, for all $ (u_ {1}, u_ {2}, ..., u_ {d}) \in
	\lbrack 0,1] ^ {d} $, by:
	\begin{equation}
	\bar{C} (u_ {1}, u_ {2}, ..., u_ {d}) = \sum_{M \subset N} (- 1)^{m} C \big (%
	(1-u_ {1}) ^ {1_ {1} \in M}, (1-u_ {2}) ^ {1_ {2} \in M}, ..., (1-u_ {d}) ^ {1_ {d} \in M} \big)
	\end{equation}%
	
	where $ N = \{1,2, ..., d \} $, $ m = \lvert M \rvert $ is the cardinality of $ M $, and $1_ {i} \in M $ indicates that $ i $ belongs to $ M $. \newline

	It is therefore advisable not to confuse the dual copula with the survival copula.
\end{remark}

\subsubsection{Sample of Copulas for finance in this study \label{exple cop}}

\textbf{ Archimedian copulas}\newline

In the literature, there are several families of copulas and some of which are more suited to financial modeling. Archimedean copulas family includes the models of Clayton, Frank and Gumbel. These copulas have the advantage of capturing the structure of positive or negative dependence between the variables. These types of dependences are characteristics of financial variables, which justifies the use of this copula family. In terms of option pricing, for example, these copulas have been used in Cherubuni and Luciano (\cite {cheru2000}) and in Slavchev and Wilkens (\cite{slavchev2009}). \newline

\begin{table}[!h]
	\centering
	\caption{Examples of Archimediean copulas.}
	\begin{tabular}{c|c|c}
		\cmidrule(r){1-3} 
		\textbf{	Family} & \textbf{ Archimedean generator} & \textbf{Copula $C_{\theta}(u)$, $u=(u_1,..., u_d)$}\\ 
		\cmidrule(r){1-3}
		Clayton & $\psi_{\theta}(t)=\frac{ t^{-\theta}-1}{\theta} $,  $\theta > 0$  &$C_{\theta}(u)= \bigg( \sum_{i=1}^d (u_i )^{-\theta}-d+1 \bigg)^{-\frac{1}{\theta}} $ \\   \cmidrule(r){1-3}
		Frank&$\psi_{\theta}(t)=\frac{-1}{\theta}\log\big(1-(1-e^{-\theta})e^{-t}\big),$  $ \theta > 0$ & $C_{\theta}(u)= \frac{-1}{\theta}\log\bigg(1-\dfrac{\prod_{i=1}^{d}(e^{-u_i \theta -1})}{(e^{- \theta -1})^{d-1}}\bigg)$ \\ \cmidrule(r){1-3}
		Gumbel & $\psi_{\theta}(t)=(-\ln t)^{\theta}$,with $\theta \geq 1$& 	$C_{\theta}(u)= \exp\bigg( - \bigg( \sum_{i=1}^d (-\log u_i )^{\theta}\bigg)^{\frac{1}{\theta}} \bigg)$ \\
		\bottomrule
	\end{tabular}
	\label{tab }
\end{table}

\textbf{Elliptical copulas} \newline

Other types of copulas used in finance are the normal copula and the t-copula. They belong to the family of elliptical copulas which describe the dependence structure of elliptical distributions. The choice of this family is justified by the fact that elliptic distributions have long been used to model random phenomena in many fields. Despite the demonstration of the leptokurtic character of the returns of financial series, of which they have the weakness to rigorously model, their continued use in finance.

The \textit{normal copula} is also known as the Gaussian copula. Its expression is:

\begin{equation}
C_{\Sigma}(u_1, u_2, ..., u_d)=\int_{-\infty}^{N^{-1}(u_1)}...\int_{-\infty}^{N^{-1}(u_d)} \dfrac{1}{(2\pi)^{d/2}\lvert \Sigma \rvert^{1/2}} \exp\big(-\frac{1}{2} y'\Sigma^{-1}y\big) dy,
\end{equation}

where $ N^{- 1} $ is the quantile of the standard normal distribution and $ \Sigma $ is the correlation matrix. \newline

The \textit {t-copula} is the one describing the dependence of t-distributions. It has the advantage of capturing the dependency in the distribution tails more than the normal copula. Its expression is given by:

\begin{equation}
C_{v, \Sigma}(u_1, u_2, ..., u_d)= \int_{-\infty}^{t_v^{-1}(u_1)}...\int_{-\infty}^{t_v^{-1}(u_d)} \dfrac{\Gamma \big( \frac{v+d}{2} \lvert \Sigma \rvert^{1/2}\big)}{\Gamma (v/2)(v\pi)^{d/2}}\big(1+\frac{1}{v} y'\Sigma^{-1}y\big)^{\frac{v+d}{2}}dy,
\end{equation}

where $ t_v ^ {- 1} $ is the quantile of the t-distribution with $ v $ degree of freedom; $ \Gamma $ is the gamma function and $ \Sigma $ the correlation matrix.

\subsubsection {Estimation-Adequacy test of a copula}

The choice of the copula that rigorously describing multivariate statistical data requires estimation and conformity testing. There are several techniques in the literature for estimating copulas belonging to different families: parametric; semi-parametric and non-parametric. For more details on these methods, see Bouy\'{e} \cite{bouye}.

\textbf {Estimation by the IFM method} \\

The IFM method (inference functions of margins) is a two-step estimation method of a copula. It was presented by Shih and Louis \cite {shih1995} in the bivariate case and then developed in dimension greater than two by Joe and Xu \cite {joe1996}. It is carried out as follows:

\begin{itemize}
	\item[1)] the first step consists in finding the estimators $ \hat{\alpha}_i $ of the parameters $ \alpha_i, \; i = 1, .., d $ for marginal distributions by maximum likelihood:
	
	$$ \hat{\alpha} _i = argmax \sum_ {j = 1} ^ {N} \ln f_n (x ^ j_i; \alpha_i); $$
	\item[2)] Once the marginal have been determined, we estimate the parameter $ \theta $ of the copula that best describes these marginal by the maximum likelihood.
	
	$$ \hat{\theta} =argmax \sum_{j=1}^{N}\ln c(F_1(x^j_1; \hat{\alpha}_1);...;F_d(x^j_d; \hat{\alpha}_d); \theta). $$
\end{itemize}

One of the advantages of this method is that under certain conditions of regularity, the IFM estimator is consistent and asymptotically normal.

Also, in terms of numerical computation time, this method is better than the "direct" maximum likelihood method since it is simpler and faster. \\

\textbf {Fit test} \newline

To confirm whether the chosen parametric copula models the data well, it is necessary to perform a test. 
The most powerful tests are based on the processes $ \sqrt {n} (\hat {C}-C_{\hat{\theta}}) $, where $ \hat{C} $ and $ C _ {\hat { \theta}} $ are respectively the empirical copula and the parametric copula.\newline

The Cramer-von Mises statistic is by far the most used because it gives satisfactory results. It is defined by:

\begin{equation}
\int_{[0;1]^d}n(\hat{C}-C_{\hat{\theta}})d\hat{C}.
\end{equation}

Others criteria such as the Akaike Criterion (AIC) and the Bayeian Inference Criterion(BIC) are very often used for the choice of the best copula. They are respectively defined by:

\begin{eqnarray}
AIC = 2m-log (l (\theta)) \\
BIC = m * log (n) -log (l (\theta)).
\end{eqnarray}
where $ l (\theta) $ is the model likelihood for the estimated parameter $ \theta $, $ m $ the number of estimated parameters and $n$ the data size.

\section{Methodology and Application}

The price of a multivariate option is a function of the density of the joint distribution. Thus, their valuation requires the determination of the joint risk-neutral density. To do this, it suffices to determine the marginal risk-neutral densities and then to choose the copula that best describes their dependence structure by using Sklar's theorem. This perspective is possible because the objective copula can be matched with the risk-neutral joint copula, under certain conditions (see Rosenbergh \cite{rosb2003}).

\subsection{Methodology}

Our approach consists firstly in determining the marginal risk-neutral distributions by the procedure used by Idier and al. \cite{jul2008}. This in order to take into account the effect of extreme values in the margins. We will also limit ourselves to the case of a mixture of two Gaussians in this study. Then, we will choose among the families of copulas listed in the section \ref{exple cop} the one that best suits the study. And finally we will determine the prices of the multivariate options by numerical integration (Monte-Carlo method) by using the formulas provided below for multivariate options considered. \newline

We will be particularly interested by rainbow options (those relating to the maximum or the minimum of several assets, etc.). These kinds of options have been the subject of many studies as in Stulz \cite{stulz1982} and Jonshon \cite{johnson1987}. \newline 

Consider $d$ assets whose price at maturity $T$ are denoted by $S_1^T; ... S_d^T$ and denote by $X_1 ^ T; ...; X_d^T$, respectively, the returns associated to each asset at  instant $T$ (with for all $ i = 1, ..., d $; $ X_i^{T}=\ ln (S_i ^{T} / S_i^t) $). For a chosen strike price $K$, we consider the following different types of rainbows: spread option; option on the maximum; option on the minimum and digital option. \newline

\subsubsection{Spread Option}

Having a pay-off equal to $ max (S_2^ T-S_1^T-K; 0) $, its value is calculated by:

\begin{eqnarray}
V_{OS}(t)&=& e^{-r(T-t)}\mathbb{E}^* \bigg\{max(S_2^T-S_1^T-K; 0)\bigg\} =e^{-r(T-t)}\mathbb{E}^* \bigg\{\int_{\mathbb{R}}\mathbb{I}_{\{K+S_1^T \leq x \leq S_2^T\}}dx\bigg\} \nonumber \\
&=& e^{-r(T-t)}\mathbb{E}^* \bigg\{\int_{\mathbb{R}}\mathbb{I}_{\{K+S_1^T \leq x \}} -\mathbb{I}_{\{K+S_1^T \leq x \; et \; S_2^T \leq x\}}dx\bigg\} \nonumber
\end{eqnarray}	

which gives,
\begin{eqnarray}
V_{OS}(t)&=& e^{-r(T-t)}\int_{-\infty}^{+\infty}\mathbb{P}^*{\{K+S_1^T \leq x \}} -\mathbb{P}^*{\{K+S_1^T \leq x \; et \; S_2^T \leq x\}}dx \nonumber \\
&=& e^{-r(T-t)}\int_{-\infty}^{+\infty}\mathbb{P}^*{\{\dfrac{S_1^T}{S_1^t} \leq \dfrac{x -K}{S_1^t} \}} -\mathbb{P}^*{\{\dfrac{S_1^T}{S_1^t} \leq \dfrac{x -K}{S_1^t}\; et \;\dfrac{S_2^T}{S_2^t} \leq \dfrac{x}{S_2^t}\}}dx \nonumber 
\end{eqnarray}	
and finaly,

\begin{eqnarray}
V_{OS}(t)&=& e^{-r(T-t)}\int_{-\infty}^{+\infty}\mathbb{P}^*{\{X_1^T \leq \log(\dfrac{x -K}{S_1^t})  \}} -\mathbb{P}^*{\{X_1^T \leq \log(\dfrac{x -K}{S_1^t}) \; et \; X_2^T \leq \log(\dfrac{x}{S_2^t})\}}dx \nonumber \\
&=& e^{-r(T-t)}\int_{K}^{+\infty} e^{r(T-t)}\dfrac{\partial  P_{t,1}}{\partial k_1}(T; \log(x_1 -k_1)) - \nonumber \\
& -& C\big( e^{r(T-t)}\dfrac{\partial  P_{t,1}}{\partial k_1}(T; \log(x_1 -k_1)) ; e^{r(T-t)}\dfrac{\partial  P_{t,2}}{\partial k_2}(T; \log(x_2))\big)dx,
\end{eqnarray}		

where 
$$x_i=\dfrac{x}{S_i^t};\; k_i=\dfrac{K}{S_i^t},$$
 and $P_{t,i}$ is the put $i$ price’s, for $i=1;2.$ \newline

\subsubsection{Call on the maximum}

Its pay-off is equal to $ max \{max (S_1^{T}; ...; S_d^{T}) - K; 0 \} $. Thus, its price at maturity is given by:

\begin{eqnarray}
V_{CMax}(t)&=&e^{-r(T-t)} \mathbb{E}^*\{max\{max(S_1^{T};...;S_d^{T})-K; 0\}\} = e^{-r(T-t)} \mathbb{E}^*\bigg\{\int_{\mathbb{R}} \mathbb{I}_{\{k \ \leq x \leq max(S_1^{T};...;S_d^{T})\}}dx \bigg\} \nonumber \\
&=& e^{-r(T-t)} \mathbb{E}^*\bigg\{\int_{K}^{+\infty} 1-\mathbb{I}_{ \{max(S_1^{T};...;S_d^{T}) \leq x \}}dx \bigg\}.  \nonumber		
\end{eqnarray}
we obtain,
\begin{eqnarray}
V_{CMax}(t)&=& e^{-r(T-t)} \int_{K}^{+\infty} 1-\mathbb{P^*}\{S_1^{T}\leq x;...;S_d^{T} \leq x \}dx  \nonumber\\	
&=& e^{-r(T-t)} \int_{K}^{+\infty} 1-\mathbb{P^*}\{X_1^{T}\leq x_1;...;X_d^{T} \leq x_d \}dx  \nonumber
\end{eqnarray}
and finaly,
\begin{eqnarray}
V_{CMax}(t)&=& e^{-r(T-t)} \int_{K}^{+\infty} 1-C\bigg(e^{r(T-t)}\dfrac{\partial  P_{t,1}}{\partial k_1}(T; x_1);...;e^{r(T-t)}\dfrac{\partial  P_{t,d}}{\partial k_d}(T; x_d) \bigg)dx \hspace{1cm}	
\end{eqnarray}

where $x_i=log(\dfrac{x}{S_i^t})$ and $P_{t,i}$ is the put $i$ price’s, for $i=1;2.$ \newline

\subsubsection{Call on the minimum}

It admits for pay-off $ max \{min (S_1^T; ... S_d^T) -K; 0 \} $ and its value at maturity is then defined by:
\begin{eqnarray}
V_{CMin}(t)&=&e^{-r(T-t)} \mathbb{E}^*\{max\{min(S_1^{T};...;S_d^{T})-K; 0\}\} 
= e^{-r(T-t)} \mathbb{E}^*\bigg\{\int_{\mathbb{R}} \mathbb{I}_{\{k \ \leq x \leq min(S_1^{T};...;S_d^{T})\}}dx \bigg\} \nonumber 
\end{eqnarray}
which is equal to,
\begin{eqnarray}
V_{CMin}(t)	&=& e^{-r(T-t)} \mathbb{E}^*\bigg\{\int_{K}^{+\infty} \mathbb{I}_{ \{min(S_1^{T};...;S_d^{T}) \geq x \}}dx \bigg\} 
= e^{-r(T-t)} \int_{K}^{+\infty} \mathbb{P^*}\{S_1^{T}\geq x;...;S_d^{T} \geq x \}dx  \nonumber	           	
\end{eqnarray}
at the end, we obtain
\begin{eqnarray}
V_{CMin}(t)	&=& e^{-r(T-t)} \int_{K}^{+\infty} \mathbb{P^*}\{X_1^{T}\geq x_1;...;X_d^{T} \geq x_d \}dx  \nonumber\\	
&=& e^{-r(T-t)} \int_{K}^{+\infty} \bar{C}\bigg(-e^{r(T-t)}\dfrac{\partial  C_{t,1}}{\partial k_1}(T; x_1);...;-e^{r(T-t)}\dfrac{\partial  C_{t,d}}{\partial k_d}(T; x_d)\bigg)dx \hspace{1cm}
\end{eqnarray}
where $x_i=log(\dfrac{x}{S_i^t})$ and $C_{t,i}$ is the call $i$ price’s, for $i=1;2.$ \newline									

\subsubsection{ Digital Option}

It has for pay-off $ \mathbb{I}_{\{S_1 ^ T \geq K_1; ...; S_d ^ T \geq K_d \}} $. Thus, its value at maturity is given by

\begin{eqnarray}
V_{ODig}(t) &=& e^{-r(T-t)}\mathbb{E}^*\big\{ \mathbb{I}_{\{S_1^T \geq K_1;...; S_d^T \geq K_d\}}\big\} =  e^{-r(T-t)} \mathbb{P}^* \big\{S_1^T \geq K_1;...; S_d^T \geq K_d\big\} \nonumber
\end{eqnarray} 
which give us,
\begin{equation}
V_{ODig}(t)	= e^{-r(T-t)} \bar{C}\bigg(-e^{r(T-t)}\dfrac{\partial  C_{t,1}}{\partial k_1}(T; k_1);...;-e^{r(T-t)}\dfrac{\partial  C_{t,d}}{\partial k_d}(T; k_d) \bigg)
\end{equation}
where   $k_i=\dfrac{K_i}{S_i^t}$; $C_{t,i}$ is the call $i$ price’s, for $ i=1;...;d$ and $\bar{C}$ the survival copula.\newline

\begin{remark}	
	Not to forget that the quantities $-e^{r(T-t)}\dfrac{\partial C_{t,d}}{\partial k}(T; k) $ and $e^{r(T-t)}\dfrac{\partial  P_{t,d}}{\partial k}(T; k) $  are both equal to the risk neutral distribution wich density is given by relation (\ref{risque-neutral density}) for our study.									
\end{remark}

\subsection{ Applications}

We will focus on the bivariate options on the pair of \textit{Atos} and \textit{Dassault Systems} shares. The data were obtained from \textit{Investing.com} and relate to the components of the \textit{CAC40 index} of the Paris stock exchange. The collected data concerns the closing prices for the period from July 01, 2014 to June 30, 2020 (1534 days). \newline

At first, for each of the two assets (Atos; Dassault Systems) the parameters of the two Gaussian regimes constituting the Gaussian mixture are determined (\textit{table \ref{tab parametre}}) as well as the proportions of each diet.\newline

\begin{table}[!ht]
		\caption {Estimated parameters of the Gaussian mixture and their proportion.}
	\begin{tabular}{llllllllllll}
		\cmidrule(r){3-6}
		\multicolumn{2}{}{}& &&  &   \\
		\multicolumn{2}{}{}&Regime 1 &Regime 2& Gaussien mixture & Empirical  \\	
		\multicolumn{2}{}{}& &&  &  distribution \\	
		\cmidrule(r){1-6}
		\multirow{5}{*}& Moyenne&   -0.0072328 &0.000764489 &0.00013704 & 0.0001370  \\ 
		&Ecart-type & 0.0603574 &0.013530408&  0.02142835 & 0.0214349\\ 
	Atos	& Skewness& 0& 0&-0.6131518& -2.823688 \\ 
		&Kurtosis&3&3&15.7&42.81632 \\ 
		& Proportion& 0.07845771& 0.921542310 & --&--\\ 
		\cmidrule(r){1-6}
		\multirow{5}{*}& Moyenne& 0.00110506&-0.00101651 & 0.0007598786&  0.0007599\\ 
		&Ecart-type & 0.01014017&0.03315738 & 0.01577694& 0.01630201\\ 
	Dassault systems	& Skewness& 0&0 &-0.03745108 & 0.1949504 \\ 
		&Kurtosis&3&3&10.00661&12.39682 \\ 
		& Proportion& 0.83729906&0.16270094 &-- &--	\\ 
		\bottomrule
	\end{tabular}
	
 \label{tab parametre}
\end{table}	

The next step, we determine the parameters $ (\alpha; \beta) $ of the stochastic discount factor defined by the relation (\ref{fact}) thanks to the assumptions of the model in particular that of lack of arbitration opportunity. The results obtained for a risk-free rate $ r = 0.025 $ are presented in the \textit{table \ref{fact par}}.

\begin{table}[!h]
		\centering
		\caption {Stochastic discount factor parameters for data returns with a risk-free rate $ r = $ 0.025.}
	\begin{tabular}{lll}
		\cmidrule(r){2-3}
		\multicolumn{1}{}{}&Atos& Dassault  \\
		\cmidrule(r){1-3}
		$\alpha$& 36.1209027 & 37.70565 \\
		$\beta$&-0.3610132 & -0.3500945 \\ 
		\bottomrule
	\end{tabular}
	\label{fact par}
\end{table}

\subsubsection{ Copulas fitting results}

We present the results of the copula estimates associated with our data. In each case, we will base ourselves on the Cramer-Von Mises statistic and/or the AIC criteria for the choice of the best copula. \newline

In the \textit{table \ref{par cop1}}, we present the estimated parameters (and their Cramer-Von statistics) for bivariate copulas. It then emerges that the three copulas with the best Cramer-Von Mises statistic are Tawn's copula, Frank's copula, Gumbel's copula in that order. \newline

The \textit{table \ref{par cop11}} gives the AIC and BIC of the parameters estimated for the bivariate copulas chosen. Based on these criteria, the four best candidate copulas for our bivariate data are the normal copula, the Husler-Reiss's copula, the Galambos's copula and the Gumbel's copula. \newline

We can notice that for our data, the performance of each fitted copula differs according to the criterion. It is then difficult to make a particular choice on a copula in suck situation on the basis of the two criteria (Cramer-Von Mises statistic vs AIC) combined. Nevertheless, if there is a choice to be made between these two criteria, it would be more judicious to base oneself on the Cramer-Von Mises test. \newline

\begin{table}[!htbp]
		\centering
		\caption {Parameters of bivariate copulas selected  and their Cramer-Von Mises test statistics.}
	\begin{tabular}{llllllll}
	\cmidrule(r){2-8}
		\multicolumn{1}{}{}&Normale & Clayton & Gumbel & Frank& Tawn& Galambos  & Husler-Reiss   \\
		\cmidrule(r){1-8}
		parameter&0.2822    & 0.1766&1.344 & 2.3166&0.6868 & 0.5995 & 0.9798 \\ 
		Statistique of Test&1.469   & 3.257	&  0.72485 	&0.6136&0.6136 &0.79162 & 0.84238	 \\ 
		p-value&0.0005  &0.0005 &0.0005 & 0.0005& 0.0005&0.0005 &0.0005   \\ 
	\end{tabular}

	\label{par cop1}
\end{table}

\begin{table}[!h]
		\centering
		\caption {AIC and BIC of the estimated parameters (by the maximum likelihood.) for the selected bivariate copulas. }
	\begin{tabular}{llllllll}
		\cmidrule(r){2-8}
		\multicolumn{1}{}{}&Normale & Clayton & Gumbel & Frank & Tawn& Galambos  & Husler-Reiss  \\ 
		\cmidrule(r){1-8}
		$log(l(\theta))$&365.9  & 325.1 & 345.4& 291.8&320 &354.7 & 358\\
		AIC& -729.8  & -648.2 &-688.8 & -581.6& -638& -707.4& -714\\ 
		BIC&-724.46 &  -642.86	& -683.46 & -576.26&-632.66 &  -702.06& -708.66\\ 
	\end{tabular}
\label{par cop11}
\end{table}

\subsubsection{ Options prices by Monte-Carlo approach}

In this section, we give the simulation results of the prices (for one period  $T=t+1$) of all options presented in the section above based on the basket (Atos, Dassault systemes). The tables \textit{\ref{prix callmax biv}} and \textit{\ref{prix callmin biv}} give, respectively,  the prices of the call on maximum and the call of maximum. We fixe the price of each asset  of the bivariate basket to $120$. The values of their prices are calculated when it is out-of-the money (OTM); at-the-money (ATM) and in-the-money (ITM). For the cases of  digital option and the spread option, we fixe, respectively, the price of the basket to $S=(120, 130)$  and $S=(100, 120)$. We give the prices of these options in the table \textit{\ref{prix option digital biv}} and \textit{\ref{prix spreadoption biv}} obtained also for different strikes. The prices are calculated  by Monte-carlo method with $N=10^5$ simulations.\footnote{The choice of the bivariate options with our underlying is simply for academic interest. In fact, they are not exchanged on the market.}\newline

For the case of the call on maximum (table \ref{prix callmax biv}), the price obtained by normale is superior than the prices with all others copulas (archimedian and extreme) in all the three situations without only the case when it is at the money with Cayton's copula. We notice that the prices obtained with the others are approximatily the same (weak discrepencies). The normale copula over-estimate the price  compared to the Tawn copula with has the good fitness test. \newline
\begin{table}[!htbp]
	\centering
	\caption{ Prices of the call on maximum with different strikes.}
	\begin{tabular}{llllllll}
		\cmidrule(r){2-8}
		\multicolumn{1}{}{}&Normale & Clayton & Gumbel & Frank & Galambos& Tawn& Husler-Reiss\\ 
		\cmidrule(r){1-8}
		OTM (K=130) &2.7126  &2.6947 &2.6642  & 2.6888 & 2.6593&2.6664 & 2.6617\\ 
		ATM(K=120)&7.328  & 7.417 &7.0226 & 7.101 &7.0216 &7.0264 &7.0245\\ 
		ITM(K=110) &17.211  &17.057 &16.743 & 16.794&16.674 &16.660	&16.673\\ 
	\end{tabular}
	 \label{prix callmax biv}
\end{table}	
 
 In the case of the call on minimum (table\ref{prix callmin biv}), the normale copula present a price which is lower than that obtained by any extreme value copula when  it is at-the-money or out-of-the money. We notice the contrary when it is in-the-money. \\
 Particularly, the normale copula give the highest price that of Tawn's copula when it is in-the-money with a small discrepency. And, when it is at-the-money or out-of-the money the Tawn's copula produice a price superior to that of normale copula with a fairly large gap than the first situation.\newline
\begin{table}[!htbp]
	\centering
		\caption{ Prices of the call on minimum with different strikes K.}
	\begin{tabular}{llllllll}
		\cmidrule(r){2-8}
		\multicolumn{1}{}{}&Normale & Clayton & Gumbel & Frank & Galambos& Tawn& Husler-Reiss\\ 
			\cmidrule(r){1-8}
		OTM(K=130)& 0.0386  &  0.02188  &  0.05882&0.03639  &0.06085 & 0.05437
		& 0.06224\\ 
		ATM(K=120) &1.181  &0.995  &1.361 &1.268  &1.356 & 1.369&1.353\\
		ITM(K=110)&10.628  &10.183 &10.559 &10.451 &10.536 &	10.523&10.543 \\ 
	\end{tabular}
 \label{prix callmin biv}
\end{table}

For prices of the digital option (table \ref{prix option digital biv}), that obtained by  normale copula is inferior to the others calculated by extreme copula in the three situations of valuation.  \newline
\begin{table}[!htbp]
	\centering
	\caption{ Prices of the bivariate digital option (paying one unit of the money) with different strikes $K$.}
	\begin{tabular}{llllllll}
		\cmidrule(r){2-8}
		\multicolumn{1}{}{}&Normale & Clayton & Gumbel & Frank & Galambos& Tawn& Husler-Reiss\\ 
		\cmidrule(r){1-8} 
		OTM&0.02283713  &0.014745475 &0.03258418 &0.02359763 &0.03306132 &  0.03181797& 0.03329281\\ 
		ATM&0.5208968  & 0.507406 &0.5273577 & 0.5344094 &0.5258449 & 0.53006026&0.5247992\\ 
		ITM&0.97519641  &0.9751979 &0.9751963 &0.9751963 &0.9751963 &0.9751964 &0.9751963\\ 
	\end{tabular}
	 \label{prix option digital biv}
\end{table}	

For the spreads option (table \ref{prix spreadoption biv}), we notice that when it is at-the-money the normale copula give the high price and the Tawn copula has the small price. Other wise, when its in-the-money the normale copula give the smallest price. Finaly, when the option is out-of-the money, the normale copula give the second greatest price. 
Comparatively to the price obtained with Tawn's copula, the price calculated with normale copula is greatest when the option is at-the-money and out-of-the-money but smallest when it is in-the-money.

\begin{table}[!htbp]
	\centering
		\caption{ Prices of the bivariate spread option  with different strikes K.}
	\begin{tabular}{llllllll}
		\cmidrule(r){2-7}
		\multicolumn{1}{}{}&Normale & Clayton & Gumbel & Frank & Galambos& Tawn\\ 
		\cmidrule(r){1-7}
		OTM(K=30) &0.04626 &0.05037 &0.01244 &0.02615 & 0.00898& 0.01909 \\ 
		ATM (K=20)&2.1429 &1.379 &0.9609 &1.0818 &1.006 &  0.8878 \\ 
		ITM (K=10)&6.5442 &7.8945 &7.4151 &7.8331 &7.5363 & 7.4094 \\ 
	\end{tabular}
 \label{prix spreadoption biv}
\end{table}	

\begin{remark}
	When $X$ and $Y$ are two random variables that modeling the returns of two shares, having a extreme value copula, one can compute the discordance function for more information about the dependance between these variables.  For more details on this measure, see Simplice and al. \cite{simplice 2009}.
\end{remark}
\section{Conclusion and discussions}

This paper proposes an approach that allows  to take in amount the effect of extreme values in the marginal and joint distribution of the underlying for the valorisation of multivariate options. For doing so, at first,  each marginal distribution of the returns of any underlying asset is modelled by a mixture of gaussiens as in Idier et al.\cite{jul2008} and the dependence structure is modelled by a  copula. The choice of the best copula is confirmed by fitting and goodness test fit. \newline 
An application is made on the basket (Atos, Dassault systems) of the  financial market CAC40 reveal that the Tawn's copula is the best for modelling the dependence structure of their returns. Thus, the prices of four type of options are calculated by use of Monte-Carlo simulation. The simulations  results show that the normale copula over-estimate the prices for the call on maximum and the spread option when they are at-the money. In the case of digital and call on minimum, this copula under-estimate the prices when the options are is at-the money.\newline

\end{document}